\input{aipcheck}

\documentclass[
    ,final            
  ]
  {aipproc}

\layoutstyle{6x9}

\def\dslash{/\kern-.1em \partial}

\def\al{\alpha}
\def\be{\beta}
\def\ga{\gamma}

\def\ep{\epsilon}

\def\ka{\kappa}
\def\la{\lambda}

\def\si{\sigma}

\def\mn{{\mu\nu}}

\def\cl{{\cal L}}

\def\prt{\partial}

\def\expect#1{\langle{#1}\rangle}

\def\half{{\textstyle{1\over 2}}}
\def\frac#1#2{{\textstyle{{#1}\over {#2}}}}

\def\lsim{\mathrel{\rlap{\lower4pt\hbox{\hskip1pt$\sim$}}
    \raise1pt\hbox{$<$}}}
\def\gsim{\mathrel{\rlap{\lower4pt\hbox{\hskip1pt$\sim$}}
    \raise1pt\hbox{$>$}}}
\def\sqr#1#2{{\vcenter{\vbox{\hrule height.#2pt
         \hbox{\vrule width.#2pt height#1pt \kern#1pt
         \vrule width.#2pt}
         \hrule height.#2pt}}}}

\def\lrDmu{\stackrel{\leftrightarrow}{D_\mu}}

\def\lrDnu{\stackrel{\leftrightarrow}{D^\nu}}

\newcommand{\beq}{\begin{displaymath}}
\newcommand{\eeq}{\end{displaymath}}
\newcommand{\bea}{\begin{eqnarray*}}
\newcommand{\eea}{\end{eqnarray*}}

\begin{document}

\title{Low Energy Tests of Lorentz and CPT Violation}

\classification{10.30.Cp}
\keywords      {Lorentz violation, CPT violation}

\author{Don Colladay}{
  address={5800 Bay Shore Road, New College of Florida}
}

\begin{abstract}
 An overview of the theoretical framework of the Standard Model Extension (SME) that allows for a parametrization of Lorentz and CPT violating effects using effective field theory will be 
presented. A review of current bounds on these Lorentz violating parameters using various low-energy tests will be reviewed. State-of-the-art measurements involving the Penning Trap, atomic clock, torsion pendulum, and resonant cavities will be discussed. Different experiments can provide stringent bounds on a variety of parameters coupled to various fundamental particles including electrons, protons, neutrons, and photons.
 \end{abstract}

\maketitle


\section{Overview of Talk}

 Bounds on Lorentz and CPT violation are typically quoted within some 
 type of theoretical formulation that allows for symmetry breaking to occur.
 A very general approach to this problem involves the Standard Model Extension (SME),
 an effective field theory that includes various Lorentz and CPT violating vectors
 and tensors coupled to standard model fields \cite{CK1:1997,CK2:1998}.  The first part of this talk will
 outline some of the key ideas involved in this construction.
 Following this, various clock-comparison, spin-polarized pendulum, penning trap,
 and resonant cavity tests will be discussed.

\section{Standard Model Extension (SME)}

CPT and Lorentz symmetry are the fundamental assumptions that go into the
construction of virtually every quantum field theory used to describe the physical world.
On the one hand, these exact symmetries should be experimentally tested to the 
fullest possible sensitivity as a matter of principle.  On the other hand, more fundamental 
theories such as quantum gravity or string theory \cite{KP1:1991,KP2:1996} 
can lead to novel effects that violate these symmetries.
In any event, it is clear that a consistent theoretical framework is required to study
possible Lorentz and CPT symmetry violations so that different experiments may be compared
with one another.

The key idea in the construction of the Standard Model Extension (SME) is that
there is some type of spontaneous symmetry breaking mechanism that leads
to a nonzero vacuum expectation value for a vector or tensor field.
The mechanism is analogous to the conventional Higgs mechanism, but uses
a vector or tensor field instead.
This naturally leads to constant background vectors and tensors that can couple
to various standard model fields.
Certain simplifying assumptions are conventionally adopted, such as only including
gauge invariant, power-counting renormalizable, and translationally invariant fields.
The resulting construction is called the minimal SME, details can be found in
the references \cite{CK1:1997, CK2:1998}.

As an explicit example, consider a term in the fundamental lagranian of the form
\beq
{\cal{L}}=\la B^\mu \overline \psi \ga^5 \ga_\mu \psi,
\eeq
involving a coupling between a vector field $B^\mu$ and a fermion field $\psi$.
If there is some mechanism present that generates a nonzero
vacuum expectation value for the vector field, say $\expect{B^\mu} = b^\mu/\la \ne 0$,
then the resulting lagrangian will contain a term
\beq
{\cal{L}} \supset b^\mu \overline \psi \ga^5 \ga^\mu \psi,
\eeq
that violates Lorentz invariance spontaneously.
This term can lead to diurnal or boost variations in energy levels of Earth-based experiments 
as the Earth rotates while the background vector field remains fixed in space.

Presumably, such correction terms should be very small to agree with current experimental
bounds.
One possible scenario is that these terms are suppressed by 
at least one power of the low-energy scale to the Planck scale: $m_l/m_{Pl} \sim 10^{-19}$.
This order of suppression may be expected if the effects are arising from some 
unknown Planck-scale physics effects, but there is no clear connection to
an underlying theory, so it is not possible to make a specific prediction at the
current time.
Remarkably, many of the current bounds from low-energy tests of Lorentz
and CPT symmetry are comparable, or better than this simple
estimate, as will be discussed below.

Each sector allows for specific types of terms.
For example, minimal extended QED (relevant for many low-energy tests) consists of a 
perturbation from the standard lagrangian
$$
\cl^{\rm QED}_{\rm lepton-photon} = 
\half i \overline{l}_A \ga^\mu \lrDmu l_A
- m_A \overline{l}_A l_A
- \frac 1 4 F_{\mu\nu}F^{\mu\nu},
\label{qedleptonphoton}
$$
with covariant derivative $D_\mu \equiv \prt_\mu + i q A_\mu$ and
the field strength 
$F_{\al\be} \equiv \prt_\al A_\be - \prt_\be A_\al$
(Note that $A=1,2,3$ corresponds to $e,\mu,\tau$ states).
The allowed perturbative terms consistent with minimal assumptions are 
conveniently classified according to their CPT properties.
The CPT-even terms involving the lepton fields are
\bea
\cl^{\rm CPT-even}_{\rm lepton} &=& 
- \half (H_l)_{\mu\nu AB} \overline{l}_A \si^{\mu\nu} l_B
\nonumber\\ &&
+ \half i (c_l)_{\mu\nu AB} \overline{l}_A \ga^{\mu} \lrDnu l_B
\nonumber\\ &&
+ \half i (d_l)_{\mu\nu AB} \overline{l}_A \ga_5 \ga^\mu \lrDnu l_B
\label{lorviolqed}
.
\eea
The CPT-odd terms involving the lepton fields are
\beq
\cl^{\rm CPT-odd}_{\rm lepton} = 
- (a_l)_{\mu AB} \overline{l}_A \ga^{\mu} l_B
- (b_l)_{\mu AB} \overline{l}_A \ga_5 \ga^{\mu} l_B.
\label{cptviolqed}
\eeq
Note that the above couplings are matrices in flavour space,
allowing for independent couplings to different flavour particles.  
This means that bounds placed on the electron sector have no
direct implications for parameters in the neutron sector, for example,
meaning that each species of particle requires independent tests.

In the pure-photon sector,
there is one CPT-even Lorentz-violating term given by
\beq
\cl^{\rm CPT-even}_{\rm photon} =
-\frac 1 4 (k_F)_{\ka\la\mu\nu} F^{\ka\la}F^{\mu\nu} ,
\label{lorqed}
\eeq
where $(k_F)_{\ka\la\mu\nu}$ is a dimensionless coupling with 
the same symmetries as the Riemann tensor.
There is also a CPT-odd pure-photon term: 
\beq
\cl^{\rm CPT-odd}_{\rm photon} =
+ \half (k_{AF})^\ka \ep_{\ka\la\mu\nu} A^\la F^{\mu\nu} 
\label{cptqed},
\eeq
where the coupling coefficient
$(k_{AF})^\ka$
is real and has dimensions of mass.
Similar corrections exist in other sectors.

There is a deep connection between CPT and lorentz invariance.
The classic CPT theorem states that under mild technical assumptions, 
any lorentz invariant, local quantum field theory with a hermitian hamiltonian
must have CPT symmetry \cite{L:1954,P:1954}.
More recently, it was demonstrated that CPT violation implies a breaking
of lorentz symmetry in any theory involving nontrivial interactions 
\cite{OG:2002}.
As a result of these theorems, it is in fact not possible to violate CPT symmetry 
in conventional quantum field theory while maintaining lorentz invariance
as is often done by directly setting the mass of particles and antiparticles to 
be different, $m \ne \overline m$, for example.

In order to apply the SME to low-energy physics, it is necessary to perform a 
nonrelativistic reduction of the hamiltonian.
A Foldy-Wouthuysen procedure can be used to do this,
a full result is given in the literature \cite{KLA:1999}, a partial result giving a flavour
for the structure is the effective particle hamiltonian
\bea
h = & m & + {p^2 \over 2 m} \\
& + & ((a_0) - m c_{00}) + (-b_j + m d_{j0}+ \half \epsilon_{jkl} H_{kl})\si^j \\
& + & \left[ -a_j + m (c_{oj}+ c_{jo})\right]{p_j \over m} + \cdots .
\eea
The corresponding antiparticle hamiltonian can be found by replacing couplings with
CPT-conjugates
($a_\mu \rightarrow - a_\mu$, $c_{\mn} \rightarrow c_{\mn}$, etc...).
Experimental bounds are (usually) found to depend on combinations of these
parameters that appear together in $h$
For example, the torsion pendulum experiments are sensitive to
the combinations
$$
\tilde b_J = b_J - \frac{1}{2} \epsilon_{JKL} H_{KL} - m d_{JT} ,
$$
where $J$ is a spatial index and $T$ is a time index in a sun-centered standard
inertial reference frame \cite{KM:2008}, often chosen as the standard reference frame 
for experiments.

\section{Low-Energy Experiments}

Clock comparison tests place stringent bounds on parameters associated
with the nuclear particles involved.  
For example, a recent experiment using a K-$^3$He co-magnetometer
\cite{BSKR:2010} placed a bound of
$$
|\tilde b^n| < 10^{-33} {\rm GeV}
$$
on neutron Lorentz violation.
The comagnetometer functions by using one atomic species (He) to shield any stray
magnetic fields while the other species senses any lorentz-violating effects on the energy
levels of the system.  
The experiment rotates with the Earth and can be independently rotated about the vertical
axis on a turntable.
This makes the experiment sensitive to several additional components of the lorentz-violating
background fields than in the earlier fixed-base version.

Spin-polarized torsion pendula are also extremely useful for low-energy tests of lorentz 
symmetry. 
The basic construction involves orienting layers of differing materials with alternate
magnetic fields that cancel out, but with an overall nonzero net spin polarization.  
The pendulum is constructed to have a large spin angular momentum,
corresponding to roughly $10^{23}$ aligned electron spins with very little
associated magnetic field.
Using this apparatus\cite{HACCSS:2008}, 
bounds on electron lorentz violation have been placed at
the level
\beq
\tilde b_j^e<10^{-31} \rm{GeV}.
\eeq
The experiment is also sensitive to boost-violating parameters 
\beq
\tilde b_T^e + \tilde d^e_- + \cdots < 10^{-27} \rm{GeV},
\eeq
at a level suppressed
by $v/c \sim 10^{-4}$, where $v$ is the velocity of the Earth.
In fact, the experiment is sufficiently sensitive that the gyroscopic coupling
between the angular momentum of the electrons and the rotation of the
Earth is a significant background signal.

Penning traps have also been a useful tool in low-energy tests of Lorentz violation.
They have a bit less sensitivity than the spin pendulum experiments, but they have the
advantage that they are sensitive to specific CPT-violating parameters,
that are mixed into more general linear combinations in other experiments.
The main feature that allows for this is the capacity to compare anomaly 
frequencies of electrons with positrons which makes it specifically a CPT test.
Bounds on the electron CPT-violating spin couplings from Penning Traps 
\cite{DMVS:1999} are at a level
\beq
|\vec b^e|<10^{-24} \rm{GeV}.
\eeq
Note that this is a very clean CPT test, rather than a bound on $\tilde b$ found using other
particle experiments.

An useful test in the photon sector involves using microwave and optical 
cavities to search for asymmetries in photon propagation properties.  
These experiments are sensitive to effects that are similar to the original Michelson-Morley
experiment.
However, rather than searching for the presence of an aether, these experiments
are currently placing stringent bounds on Lorentz-violating photon couplings.
Birefringent terms are typically bounded at the $10^{-42}$GeV level using various astrophysical
measurements \cite{KM:2008}.
This implies that the cavity tests are particularly useful for bounding non-birefringent 
photon couplings that are not easily accessible using astrophysical experiments.
One example is the bound \cite{H:2009}
\beq
(\tilde \kappa_{e-}) < 10^{-17},
\eeq
where $(\tilde \kappa_{e-})^{JK} = - k_F^{TJTK} + \cdots$ is a linear combination of
photon lorentz-violating terms that the experiments are sensitive to.

\section{summary}
Low-energy tests of Lorentz invariance place some of the most stringent bounds
to date on certain combinations of parameters in the SME.  
Current bounds are at or even in excess of naive estimates based on Planck-scale
suppression factors.
This indicates that there is a real possibility that these experiments may see effects
that might arise from a deeper theory that allows for Lorentz or CPT breaking.
For more information, the Data Tables for Lorentz and CPT Violation \cite{R:2012} contain a comprehensive list of current bounds in all sectors within the SME framework.


\begin{theacknowledgments}
 I would like to thank New College for providing me with summer research funds that helped
 contribute to the completion of this work.
 \end{theacknowledgments}



\bibliographystyle{aipproc}   

 \bibliography{proc2012}

\IfFileExists{\jobname.bbl}{}
 {\typeout{}
  \typeout{******************************************}
  \typeout{** Please run "bibtex \jobname" to optain}
  \typeout{** the bibliography and then re-run LaTeX}
  \typeout{** twice to fix the references!}
  \typeout{******************************************}
  \typeout{}
 }

\end{document}